# Development of a Plasticity-oriented Interatomic Potential for CrFeMnNi High Entropy Alloys

Ayobami Daramola[a*], Giovanni Bonny[b], Gilles Adjanor[c], Christophe Domain[c], Ghiath Monnet[c], Anna Fraczkiewicz[a]

[a]MINES Saint-Etienne, Université de Lyon, CNRS, UMR 5307 LGF, Centre SMS, 42023, ́Saint-Etienne, France
[b]SCK CEN, Nuclear Materials Science Institute, Boeretang 200, B-2400 Mol, Belgium
[c]Electricité de France, EDF Recherche et Développement, Département Matériaux et Mécanique des Composants, Les Renardières, F-77250 Moret sur Loing, France
Corresponding author: Ayobami Daramola[a], Anna Fraczkiewicz[a]
Email address: ayobami.daramola@emse.fr fraczkie@emse.fr

**Abstract**

An interatomic potential (termed EAM-21) has been developed with the embedded atomic method (EAM) for CrFeMnNi quaternary HEAs. This potential is based on a previously developed potential for CrFeNi ternary alloys. The parameters to develop the potential were determined by fitting to experimental values, density functional theory (DFT) and thermodynamic calculations, to reproduce the main crystal characteristics, namely: the stability of the fcc phase, elastic constants, and stacking fault energy. Its applicability for the study of plastic deformation mechanisms was checked by calculations of behaviour of a $½<110>\{111\}$ edge dislocation in equiatomic quaternary CrFeMnNi alloy, as well as its less-complex subsystems (ternaries, binaries, and pure metals). The calculations were performed in the domain of temperatures between 0 and 900 K; smooth and stable glide of an edge dislocation and fcc phase stability in this temperature range was confirmed. This study demonstrates the suitability of the EAM-21 potential for the analysis of plasticity mechanisms and mechanical properties of CrFeMnNi HEAs.

# 1  Introduction

High Entropy Alloys (HEAs) are a promising group of metallic alloys, first proposed in 2004 by two almost simultaneous publications coming from the group of B. Cantor in Oxford [1] and J.W. Yeh in Taiwan [2]. These alloys generally contain at least five metallic elements present in a near-equimolar ratio: this concentrated composition is supposed to stabilize a single-phase solid solution due to the high configurational entropy of the system. Since 2004, the concepts around "HEAs" have been strongly evolved; especially, significant deviations from stoichiometry have been observed in different single-phase HEAs. Moreover, some intriguing properties that seemed to be specific to HEAs have also been observed in other concentrated alloys like CoCrNi [3], and obviously, the entropy criterion was quickly shown not to be a sufficient condition to create a concentrated solid solution [4]. However, HEAs remain one of the topics of interest in the physical metallurgy community, and numerous studies have allowed the discovery of new alloys of different structures: face-centred cubic (fcc) [1, 5, 6], body-centred cubic (bcc) [7, 8], as well as hexagonal-close packed (hcp) [9, 10]. The properties of HEAs, which are sometimes different from classical alloys, seem to come from four "core effects" proposed by Yeh et al. [2]: (i) high entropy effect; (ii) severe lattice distortion; (iii) sluggish diffusion and (iv) cocktail effect. Nevertheless, some of these effects have been subjected to questions [4], which have paved the way for future studies in HEAs.

The first proposed HEA was fcc-structured equimolar CoCrFeMnNi, currently named *Cantor alloy* [1]. It remains today the most studied alloy considered as a model structure for understanding and development of new grades of HEAs. Many alloys from the Cantor family present a set of common characteristics [3, 11-21]: stable fcc structure, with no traces of martensitic transformation even under stress, high solid solution strengthening [12], good plasticity, and reasonably high static mechanical resistance associated to a simultaneous increase of both mechanical resistance and elongation to fracture when the temperature decreases. Several detailed studies have linked this property to the tendency of fcc-structured HEAs to twin at low temperatures [3, 14, 24, 25]. Furthermore, a recent study [13] based on *in situ* TEM observations of deformation mechanisms, suggests that the twinning phenomenon in the Cantor alloy is present both at room and cryogenic temperatures, and its probability depends mostly on the stress conditions (crystal orientation).

Dislocation mechanisms in this family of HEA seem similar to what is currently observed in austenitic stainless steels (CrFeNi alloys): dislocations with Burger's vector a/2<110> glide in {111} planes. A low value of the stacking fault energy (SFE) [23] leads to a dissociation of a perfect dislocation into two partial dislocations following the Shockley mechanism: ($\frac{a}{2}[\bar{1}10] \leftrightarrow \frac{a}{6}[\bar{1}2\bar{1}] + \frac{a}{6}[\bar{2}11]$) [28]. The two partials are separated by stacking fault ribbons; in agreement with the elastic theory of dislocations, their distance is inversely proportional to the SFE value [29]. It is currently admitted that a low SFE favours twining and planar glide while high SFE values prevent twining and facilitate cross-slip of perfect dislocations [30]. It is important to note that in HEAs, SFE seems not to have a unique value due to the composition heterogeneity in the system. Therefore, local variations of SFE may lead to different local conditions of dislocation dissociation [21, 31].

As for any new material, the future use of HEAs is expected to bring significant progress in engineering materials in all demanding industries. Among them, nuclear applications are identified since several studies have shown, both by numerical simulation [26] and experimental observations [19, 20], improved resistance of Cantor alloy to irradiation damages, as compared to standard austenitic stainless steels. However, the presence of Cobalt excludes CoCrFeMnNi alloys from nuclear applications due to the production of $^{60}$Co radioisotope in in-service conditions [20, 27]. This is the main motivation for the design of new Co-free alloys, mainly from the CrFeMnNi family. In the past few years, several Co-free HEAs have been designed and tested. The $Cr_{18}FeMnNi$ (wt.%) HEA shows mechanical properties at room temperature similar to standard stainless steels (yield strength (YS) near 300 MPa, and elongation below 40 %) [27]. In the same alloy, Kumar et al., [20] show a decrease in radiation damages at elevated temperatures (500°C) as compared to CrFeNi or CrFeMn austenitic steels. The "Y3 grade", $Fe_{46}Cr_{14}Mn_{18}Ni_{22}$, designed and tested by Olszewska [19] shows increased stability of fcc structure at 500°C, as compared to Cantor alloy. Moreover, through strain hardening of this alloy, a significant increase in YS (up to 600 MPa) may be obtained without a loss of its ductility.

Further improvement of alloys from the CrFeMnNi family, both from the point of view of their mechanical resistance and in-service irradiation behaviour, needs an in-depth analysis of the atomic-scale structure of crystal defects and its elementary plastic deformation mechanisms. In this area, atomistic simulations should bring an important insight.

Among atomic-level simulation methods, the first principles (ab initio) methods carried out by density functional theory (DFT) calculations [32], take a place of choice in the investigation of materials properties. This approach leads to very reliable results but remains limited by the number of atoms and expensive computation time. To overcome this problem, the use of interatomic potentials brings an interesting compromise between computational efficiency and physical accuracy [33-35]. In this approach, the interactions between atoms in the crystal are described by potential functions that reproduce the main crystal characteristics. The potential parameters are fitted to obtain a satisfactory agreement with the real material properties, as known from existing experimental or DFT data. The main interatomic potential formalisms describing metallic systems include the embedded atom method (EAM) [36], Finnis-Sinclair method (FS) [37] and modified embedded atom method (MEAM) [38]. Over the past 35 years, these potential formalisms have been employed for atomistic simulations of various material behaviours such as phase transformations, solid solution strengthening, or dislocation motion [39]. Yet, even if the CrFeNi system, representative of currently used austenitic stainless steel, are well described, and largely studied, the Mn-containing HEA are much less known. The presence of Mn makes the development of a reliable interatomic potential for CrFeMnNi difficult, as Mn structures are complex and undergo several allotropic transformations and strong magnetic effects [35, 40]. To the best of our knowledge, currently, only three groups have developed a reliable potential for pure manganese through a MEAM approach [35, 41] or an EAM one [40].

In this paper, a new potential for CrFeMnNi has been constructed based on the EAM formalism [36]. The work takes advantage of previous results performed by employing the CrFeNi potential developed in [33] and Mn potential developed in [40]. Thus, the plasticity-oriented CrFeNi potential developed in [33] which is termed EAM-11 was used for the description of this ternary system. This potential was successfully used in [42] to investigate the interaction of an edge dislocation with Frank loops in $Cr_{20}Fe_{70}Ni_{10}$ alloy. In this work, the cross pair potential method was adopted for the addition of the Mn potential function to the existing CrFeNi potential. This concept satisfies the condition (zero pressure) involving pure potentials. The developed potential was fitted on a set of physical data (fcc structure stability, elastic constants, and stacking fault energy) and validated on its capacity to describe the structure and glide of an edge dislocation.

## 2 Methods

### 2.1 Potential development

In the EAM formalism [36] used to describe the interactions between atoms, the total energy $E$ of a system containing $N$ atoms of different species is given by Eq. 1:

$$E_i = \frac{1}{2}\sum_{j \neq i}^{N} V_{x_i x_j}(r_{ij}) + \sum_{i=1}^{N} F_{x_i}(\rho_i) \quad (1)$$

where the first term is the sum of pair potential $V_{x_i x_j}(r_{ij})$ as a function of the distance $r_{ij}$ between atom $i$ and $j$ with chemical species $x_i$ and $x_j$. The second term is the sum of embedding energies $F_{x_i}(\rho_i)$ that approximates the many-body contribution of all nearby atoms as a function of the host density $\rho_i$. The host density $\rho_i$ is expressed as the sum of contribution $\varphi_{x_j}(r_{ij})$ from all neighbouring atoms $j$:

$$\rho_i = \sum_{j \neq i}^{N} \varphi_{x_j}(r_{ij}) \quad (2)$$

The EAM-21 potential development was based on existing EAM-11 functions for CrFeNi [33] and pure Mn [40] both developed by Bonny et al.. The properties of Fe, Ni, Cr, and Mn are reported in detail [33, 40] and will not be given here. Only the Mn-involving cross-interaction functions $V_{FeMn}, V_{NiMn}, V_{CrMn}$ were fitted in this work. The FeMn, NiMn, and CrMn binary systems were benchmarked to available experimental and DFT data. The pair-interaction potential $V_{XY}$ is defined as a linear combination of piecewise cubic splines, as shown in Eq. 3.

$$V_{XY}(r) = \sum_{k=1}^{N_p} [a_k(r_k - r)^3 \Theta(r_k - r)] \quad (3)$$

$N_p$ denotes the number of knots, $r_k$ is the positions of the cubic spline knots, $a_k$ is the fitted parameters and $\Theta$ is the Heaviside step function defined as $\Theta(x) = 1$ for $x \geq 0$ and $\Theta(x) = 0$ for $x < 0$.

The potential is fitted by manual setting the knot values $r_k$ while the $a_k$ parameters are fitted to match the target values following the method described in [33, 34, 42]. To improve the accuracy of the potential, several fitting trials were carried out by fine-tuning the knot value $r_k$. This procedure was employed for all binary and ternary systems, as well as the quaternary system of interest.

The cross potentials were fitted on a set of physical properties of crystals from the CrFeMnNi family, starting with binary compositions and going to quaternary compositions. However, given the small amount of available experimental/DFT data, all data had to be included in the fitting database. To avoid overfitting and ensure maximum transferability of the potential, care was taken that the fitted cross potentials do not take unphysical shapes, e.g., strong oscillations. The set of tested properties contains the fcc structure stability, associated with the calculation of mixing enthalpy for fcc structure of all the composition range of $Cr_xFe_{1-x-y-z}Mn_yNi_z$, where x, y, z is the concentration in the range of 0 to 1. The evaluation of the elastic constants and stacking fault energy were carried out during the fitting procedure. All the calculations have been performed at 0 K. Dislocation properties have been estimated from a calculation of the stacking fault energy, for a defect resulting from a Shockley dissociation, and a simulation of glide of a dissociated dislocation. The optimized parameter set for the EAM-21 potential is given in the appendix Table A.1.

## 2.2 Simulation setup

Simulation boxes in the x, y, z directions corresponding to $20 \times 20 \times 20$ unit cells and containing 32,000 randomly distributed atoms were used for static calculations at 0 K. It should be noted that all boxes contain randomly solid solutions except for a fitting data that required the L10 intermetallic structure. Molecular dynamics calculation of dislocation behaviour was performed with a simulation cell of $200\text{Å} \times 65\text{Å} \times 133\text{Å}$ in the directions x, y, z and contained 144900 randomly distributed atoms in total. The static calculations were performed with the DYMOKA code developed by Becquart et al., [43]. For molecular dynamics (MD) dislocation calculations, LAMMPS software developed by Plimpton et al., [44] was used. OVITO software [45] and DXA (dislocation extraction algorithm) function incorporated in OVITO developed by Skukowski et al., [46] was employed for visualizing the dissociated dislocations. The first step of calculation deals with the prediction of *fcc structure stability*. In this work, the fcc phase is the ground state for Cr-Fe-Mn-Ni and its subsystems, although only Ni should have an fcc ground state at atmospheric pressure at 0 K. For Fe, Cr which have a "natural" bcc ground state and Mn which is a close-packed intermetallic structure, at atmospheric pressure and low temperature, we artificially stabilized them to have fcc phase. This was done to avoid complexities involving (local) fcc-bcc transformation with temperature and composition. For this purpose, the atoms in the simulation box are relaxed until they reach minimum energy while the pressure is maintained at a constant value, here zero. The simulated fcc structure was compared to the bcc one and the norm of the cohesive energy difference between both structures ($\Delta E_{fcc-bcc} = |E_{fcc} - E_{bcc}|$) was estimated. Moreover, the fcc *mixing enthalpy* at 0K was calculated following Eq. 4:

$$H_{mix} = E_{Cr_xFe_{1-x-y-z}Mn_yNi_z} - xE_{Cr} - yE_{Mn} - zE_{Ni} - (1-x-y-z)E_{Fe} \qquad (4)$$

where $E_{Fe_{1-x-y-z}Ni_xCr_yMn_z}$ is the energy of the weighted combined alloy and $E_{Ni}$, $E_{Cr}$, $E_{Mn}$, $E_{Fe}$ are the cohesive energies of chemical species (Ni, Cr, Mn and Fe, respectively).

For identification of the *elastic constants* $C_{ij}$ of the crystal, the simulation box was relaxed and deformed with a 0.0001 strain applied to the system, and the potential energy of the system was calculated. The three independent single-crystal constants of a cubic crystal, $C_{11}$, $C_{12}$, and $C_{44}$, the bulk modulus $B = (C_{11} + 2C_{12})/3$, tetragonal shear modulus $C' = (C_{11} - C_{12})/2$ and the isotropic shear modulus $G = (3C_{44} + 2C')/5$ were calculated. The values used in this work are calculated following the Voigt-Reuss-Hill approximation method [47]. This method takes the average between the upper bound and lower bound through the Voigt [48] and Reuss [49] models, respectively.

For the calculation of *stacking fault energy ($\gamma_{SFE}$)* at 0 K, a simulation box containing 96000 randomly distributed atoms was used, with a coordinate system in which x, y, and z axes are parallel to the crystallographic directions [110] (Burger's direction), [1$\bar{1}$2] (dislocation line) and [$\bar{1}$11] (glide plane), respectively. In this configuration, the stacking fault plane is perpendicular to the [$\bar{1}$11] (z) direction. The periodic boundary conditions were imposed in the x and y directions and a free boundary condition was applied in the z-direction. The SFE is defined as:

$$\gamma_{SFE} = \frac{E_f - E_i}{A} \qquad (5)$$

where $E_f$ and $E_i$ are the energies of the system after and before creating stacking fault, respectively, and $A$ is the area of stacking fault. It should be noted that intrinsic stacking fault was considered in this work. The $\gamma_{SFE}$ value presented in this work was obtained by averaging the calculation over 600 runs corresponding to different random atomic configurations within the calculation box. The resulting error, defined as the 95 % confidence interval around the average value, was lower than 0.3 mJ/m$^2$.

The so-obtained EAM-21 potential was tested on *dislocation movement* by a molecular dynamic (MD) calculation (LAMMPS software [44]). To introduce a ½ <110>{111} perfect edge dislocation into the crystal, the Osetsky and Bacon model [50] was adopted. The crystallographic coordinates of the simulation box were the same as mentioned before for stacking fault energy calculation, with the dislocation line following the y-axis and its Burgers vector parallel to the x-axis. The energy of the system was relaxed with a conjugate gradient algorithm that led to dislocation dissociation due to low stacking fault energy. The equilibrium dissociation width is measured using OVITO [45].

Finally, *simulations of crystal deformation were performed* by the MD method on a dislocation-containing system, in the equimolar alloy of CrFeMnNi and its subsystems. Stress was applied on the top half of the simulation cell, leading to its glide in respect to the bottom half followed by the Burgers vector (a/2[110]) direction on the ($\bar{1}$11) plane. Importantly, the NVE ensemble was applied in the simulation, where N (the number of atoms), V (the volume), and E (the total energy) of the system are conserved. A strain rate of $10^8 \ s^{-1}$ was applied on the system with a temperature from 300 K to 900 K and a time step of 1fs at each temperature. The variation of friction stress and critical resolved shear stress (CRSS) to overcome the resistance of dislocation motion was determined for all studied temperatures.

## 3 Results and discussion

### 3.1 *Stability of the fcc phase*

Table 1: The table contains minimum and maximum values predicted by EAM-21 potential for $\Delta E_{fcc-bcc}$, which is the norm of the cohesive energy difference between fcc and bcc (in eV/atom), $H_{mix}$ defined as mixing enthalpy for fcc structure (in eV/atom), and $C'$ is tetragonal shear modulus (in GPa). For comparison, the table also includes DFT and cluster expansion (CE) calculation for mixing enthalpy reported by Fedorov et al., [26].

|  | EAM-21 potential | Reference | | Composition |
|---|---|---|---|---|
| $\Delta E_{fcc-bcc}$ (eV/atom) | | | | |
| Minimum value | 0.03 | | | Fe$_{0.05}$Ni$_{0.5}$Cr$_{0.45}$ |
| Maximum value | 0.19 | | | Pure Ni |
| $H_{mix}$ (eV/atom) | | DFT | CE | |
| Minimum value | -0.18 | -0.20[a] | -0.15[a] | Ni$_{0.5}$Mn$_{0.5}$ |
| Maximum value | 0.08 | | 0.05[a] | Fe$_{0.5}$Mn$_{0.5}$ |
| $C'$ (GPa) | | | | |
| Minimum value | 15 | | | Ni$_{0.35}$Cr$_{0.65}$ |
| Maximum value | 84.4 | | | Cr$_{0.35}$Mn$_{0.65}$ |

[a] – Fedorov et al., [26]

The approach presented here aimed to make the fcc structure more stable than the bcc structure for alloys from the CrFeMnNi system in the compositions range as large as possible and at absolute zero temperature. To check this approach, a massive calculation was carried out with compositions of $Cr_xFe_{1-x-y-z}Mn_yNi_z$ (x = 0.0, 0.5, …1, y = 0.0, 0.5, …1, z = 0.0, 0.5, …1). It should be noted that obtaining fcc stability in the whole concentration range is not realistic due to the temperature effect [51]. The results obtained for the norm of the cohesive energy difference between fcc and bcc structure ($\Delta E_{fcc-bcc}$) at 0 K by EAM-21 potential are shown in Table 1. For all compositions, the minimum value gives 0.03 eV/atom for $Fe_{0.05}Ni_{0.5}Cr_{0.45}$ and the maximum value is equal to 0.19 eV/atom for pure Ni. $\Delta E_{fcc-bcc}$ is expected to be positive to attain a stable fcc structure. The fcc-bcc energy difference was not fitted to any reference functions, except to make the fcc phase as stable as possible over the complete concentration domain.

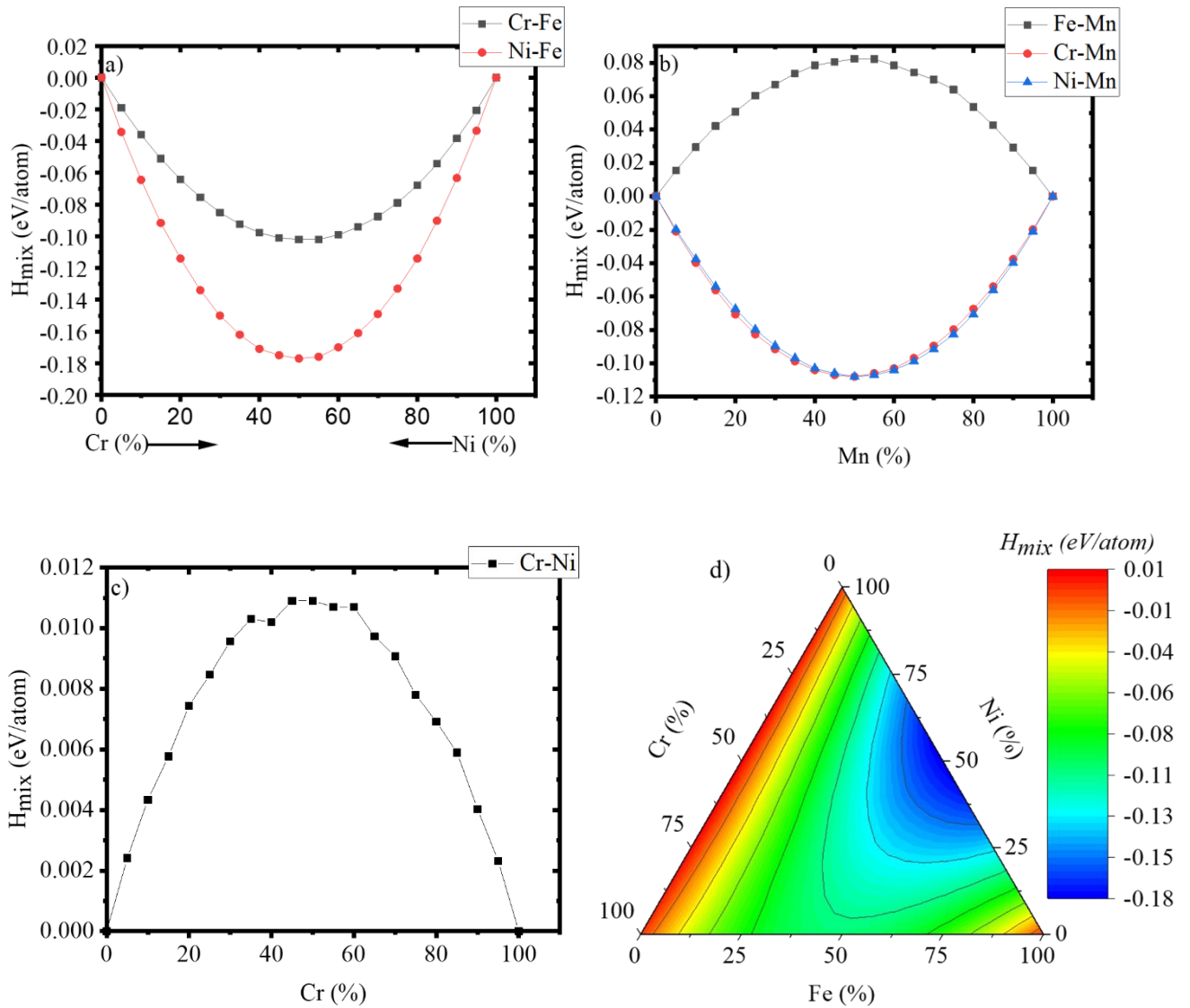

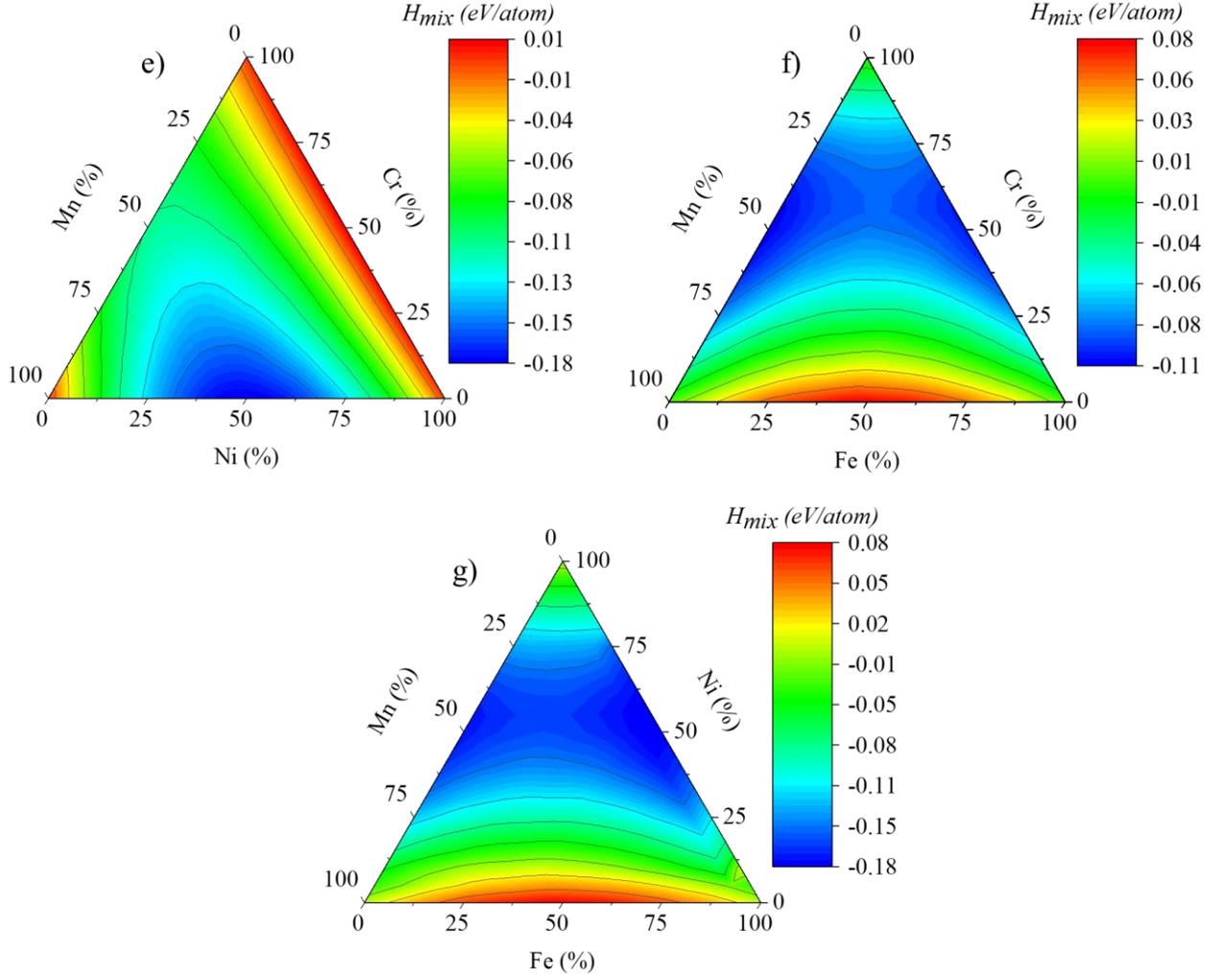

Figure 1: Enthalpy of mixing (eV/atom) of the fcc phase for CrFeMnNi at 0 K. (a-c) binary systems (Cr-Fe, Ni-Fe, Fe-Mn, Cr-Mn, Ni-Mn, and Ni-Cr), and (d-g) ternary systems (Cr-Fe-Ni, Cr-Mn-Ni, Cr-Fe-Mn, and Fe-Mn-Ni).

It is important to note that particular interest was placed on ***tetragonal shear modulus*** $C'$ which defines the stability of the fcc phase under small (elastic) deformation. For the fcc phase to remain stable under deformation, $C'$ must be positive (in practice > 10 GPa). The EAM-21 potential prediction for $C'$ was > 15 GPa for the whole concentration range of the CrFeMnNi system. The elastic constants for fcc cubic crystal should satisfy the mechanical stability criteria that are: $C_{11} - C_{12} > 0, C_{11} > 0, C_{44} > 0, C_{11} + 2C_{12} > 0$. The stability criteria require $C_{12} < C_{11}$ to obtain positive $C'$.

The ***fcc mixing enthalpy*** $H_{mix}$ values for the CrFeMnNi binary and ternary subsystems are summarized in figure 1(a-g), where figures 1(a-c) show the six binary systems (Cr-Fe, Ni-Fe, Fe-Mn, Cr-Mn, Ni-Mn, and Ni-Cr) and figures 1(d-g) corresponds to the four ternary systems (Cr-Fe-Ni, Cr-Mn-Ni, Cr-Fe-Mn, and Fe-Mn-Ni). The $H_{mix}$ values obtained from EAM-21 potential for fcc phase at 0 K gives a minimum value of -0.18 eV/atom for $Ni_{0.55}Mn_{0.45}$ and the maximum value of 0.08 eV/atom for $Fe_{0.5}Mn_{0.5}$. Let us remember that negative mixing enthalpy indicates solid solution stability and a tendency to order while positive mixing enthalpy favours phase separation. The chosen definition implies that all compositions with positive enthalpy of mixing will favour demixing and compositions with negative enthalpy mixing will show the tendency to order. Depending on the target compositions needed to be studied, the potential can reproduce the effect of either ordering or demixing, even before accounting for temperature effects (mixing entropy). The EAM-21 potential predictions are reasonably consistent with the results obtained by Fedorov et al., [26] as shown in Table 1. The authors obtained a minimum value of -0.15 eV/atom and maximum value of 0.05 eV/atom using cluster expansion calculation and minimum value of -0.20 eV/atom and maximum value of 0.05 eV/atom using DFT calculation for $Ni_{0.5}Mn_{0.5}$ and $Fe_{0.5}Mn_{0.45}$, respectively. For more details see figure S.1(a-f) in the supplementary data for the 3D plot of $\Delta E_{fcc-bcc}$, $C_{11}$, $C_{12}$, $C_{44}$, $C'$ and fcc mixing enthalpy $H_{mix}$.

## 3.2 Elastic properties

The bulk and shear modulus (respectively noted B and G) are important material properties that may be deduced from fundamental elastic constants. Their values are precious for fitting an interatomic potential focusing on mechanical properties. The graphs in figure 2 show a comparison of the calculated B and G moduli with bibliographic data. Due to the small number of available data, all kinds of information (DFT and experimental data) are used. The red line in the graphs shows correlation conditions, for which prediction by EAM-21 potential would be in perfect agreement with the existing data, while two border green dashed lines show the results in the area within the 15 % relative error. The horizontal bars show the range in between several values from the same composition is available from literature studies. Let's also note that experimental values from Teklu et al., [52] and Ledbetter et al., [53] were obtained from measurements at room temperature (295 K) while the DFT values calculated by Reeh et al., [54] and Niazi et al., [55] were obtained at 0 K. All the predicted values including $C_{11}$, $C_{12}$, $C_{44}$, $C'$ and Poisson ratio can be seen in supplementary Table S.1-S.5.

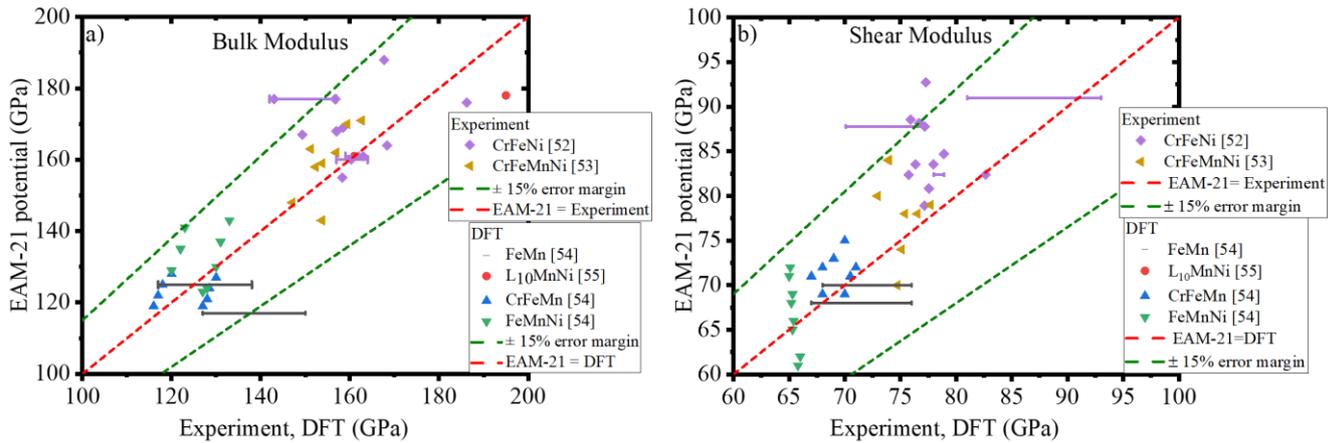

Figure 2: Elastic constants (B and G moduli) for alloys from the CrFeMnNi system calculated with EAM-21 potential and compared to the available data. The red line is a trend curve, corresponding to a perfect agreement between prediction and existing data (both experimental measurements and DFT data); the green ones give the values of +/- 15% of error margin for it. a) bulk modulus, B, and b) shear modulus, G. Calculations performed at 0 K. It is important to note that the data points are the target values used for fitting.

The EAM-21 potential predictions show an overall good agreement with experimental measurements and DFT data. The results are mostly within the 15 % relative error.

## 3.3 Stacking fault energy

The *γ*<sub>SFE</sub> is one of the basic properties to understand the deformation mechanisms of a material. It is important to obtain a reasonable *γ*<sub>SFE</sub> relative to existing data with an interatomic potential that will describe the plasticity of a material. The *γ*<sub>SFE</sub> values calculated with EAM-21potential are shown in figure 3 the influence of Mn and Ni content on the *γ*<sub>SFE</sub> in the CrFeMnNi system are shown.

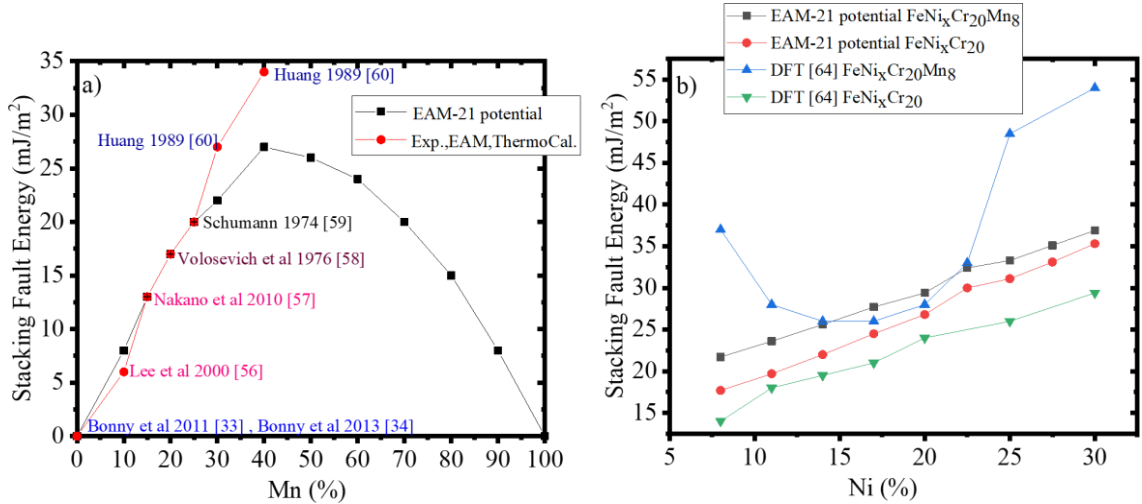

Figure 3: The effect of Mn and Ni content on the $\gamma_{SFE}$ of a) FeMn binary system and b) CrFeNiMn quaternary system.

In figure 3a, the $\gamma_{SFE}$ for FeMn binary system predicted by EAM-21 potential relative to $\gamma_{SFE}$ values obtained from experiments and thermodynamic calculations are presented [56-62]. The present calculation exhibits very good agreement up to 30 % Mn. Lee et al., [56] and Nakano et al., [57] calculated the $\gamma_{SFE}$ between 5-6 mJ/m$^2$ and 13 mJ/m$^2$ for 10 %, 15 % Mn content respectively, both authors employed thermodynamic calculation considering the chemical Gibbs energy difference between the fcc and hcp phases. Volosevich et al., [58] and Schumann et al., [59] experimentally estimate $\gamma_{SFE}$ to be 17 mJ/m$^2$ and 20 mJ/m$^2$ corresponding to 20 % and 25 % Mn content whereas Huang [60] also adopted CALPHAD model-based thermodynamic calculation to estimate $\gamma_{SFE}$ as 27 mJ/m$^2$ and 34 mJ/m$^2$ for 30 %, 40 % Mn content respectively.

In figure 3b, the effect of Ni content on the $\gamma_{SFE}$ of CrFeNi ternary and CrFeMnNi quaternary alloy is shown. The results obtained by Lu et al., [64] using DFT calculations are shown and compared with predictions by EAM-21 potential. The values predicted with EAM-21 potential show an almost linear increase in $\gamma_{SFE}$ value with increasing Ni content between 0 and 30 %, both for ternary and quaternary alloys. Moreover, a slight increase in $\gamma_{SFE}$ value is observed when Mn is added to the CrFeNi system. The comparison between EAM-21 potential and DFT calculations by Lu et al., [64] are in good agreement while the ternary system is of concern, but they diverge strongly for the Mn-containing material. The DFT calculations show a local minimum in $\gamma_{SFE}$ values, for about 17 % of Ni, because of the randomized local magnetic moment included in $\gamma_{SFE}$ calculations by Lu et al., [64]. It is important to state that the magnetic effect is not reflected in the classical many-body potential formalism which is included in DFT and CALPHAD model-based thermodynamic calculations.

Finally, figure 4 summarises all the available bibliographic data for $\gamma_{SFE}$ values and correlates it to the EAM-21 potential predictions. As most of the $\gamma_{SFE}$ values are regrouped in an area of a limited range (below 40 mJ/m²), figure 4b shows a magnification of the figure from the left, limited to the area of interest. The red line in the graphs shows the ideal agreement between prediction by EAM-21 potential and the existing data, while the two border green dashed lines show the results in the area within the 15 % relative error. The average $\gamma_{SFE}$ from existing data obtained from several values of the same composition is represented by the horizontal bar. Most of the $\gamma_{SFE}$ value predicted for the CrFeMnNi system falls within the area of 15 % margin error. Clearly, EAM-21 potential predictions show good agreement with experimental measurements, DFT and thermodynamic calculations. All the predicted values can be seen in supplementary Table S.6-S.9.

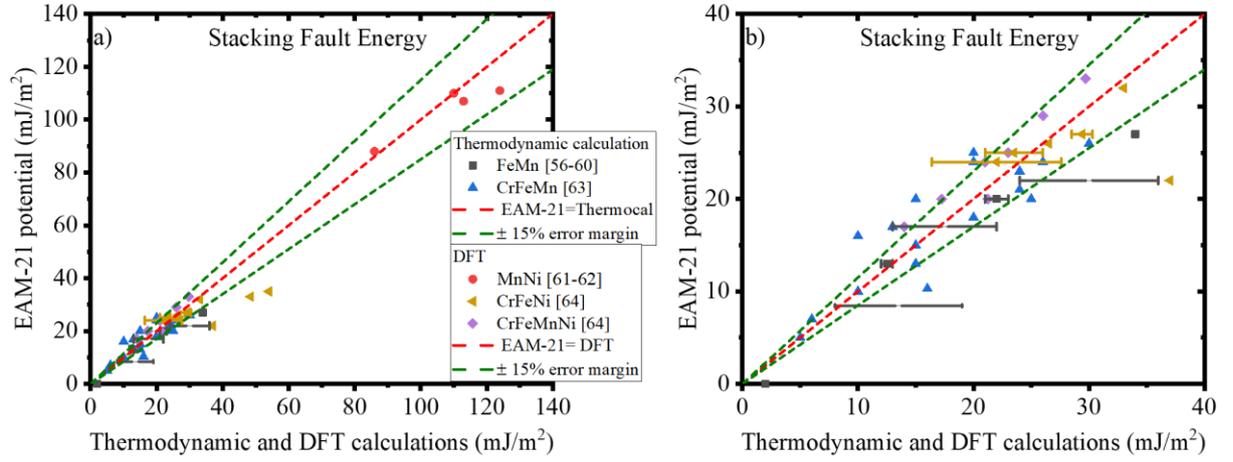

Figure 4: $\gamma_{SFE}$ for alloys from the CrFeMnNi system calculated with EAM-21 potential and compared to the available data. The red straight dot line corresponds to an ideal agreement between prediction and existing data; the green ones give the values of +/- 15% of error margin for it. a) $\gamma_{SFE}$ and b) A magnification of (a) for a range of $\gamma_{SFE}$ below 40 mJ/m$^2$.

### 3.4 Dislocation motions

As a part of the validation of the developed EAM-21 potential for the modelling of dislocation behaviour, a preliminary study on dislocation motion was carried out.

### 3.5 Equilibrium dissociation width

A glissile a/2<110>{111} edge dislocation that dissociates into two Shockley partials was studied. The calculations have been performed by MD (LAMMPS code) for the Ni-based equiatomic alloy system. The dissociation width values estimated in this work are summarized in Table 2. For this calculation, the shear modulus G is essential, it is key to understand the material deformation, so an appropriate shear modulus G value can help one to determine the equilibrium dissociation width. The shear modulus $G_H$ value reported in Table 2 is obtained by the Voigt-Reuss-Hill approximation method [44].

Table 2. Parameters used in calculating dissociation distance with elasticity theory. Estimation of equilibrium dissociation width at 0 K. Stacking fault energy $\gamma_{SFE}$, shear modulus $G_H$, Poisson ratio $\nu_H$ given by Hill average and dissociation distance $d^{EAM-21}$ predicted by the EAM-21 potential.

| Composition | $\gamma_{SFE}$ (mJ/m$^2$) | $G_H$ (GPa) | $\nu_H$ | $d^{EAM-21}$ (nm) |
|---|---|---|---|---|
| Ni | 113 | 87 | 0.29 | 1.8 |
| Cr$_{50}$Ni | 18 | 50 | 0.33 | 7.1 |
| Mn$_{50}$Ni | 38 | 81 | 0.29 | 5.5 |
| Fe$_{50}$Ni | 56 | 117 | 0.28 | 5 |
| Cr$_{33}$Mn$_{33}$Ni | 38 | 78 | 0.30 | 5.8 |
| Cr$_{33}$Fe$_{33}$Ni | 46 | 86 | 0.30 | 5.5 |
| Fe$_{33}$Mn$_{33}$Ni | 35 | 95 | 0.28 | 5.7 |
| Cr$_{25}$Fe$_{25}$Mn$_{25}$Ni | 37 | 88 | 0.29 | 5.8 |

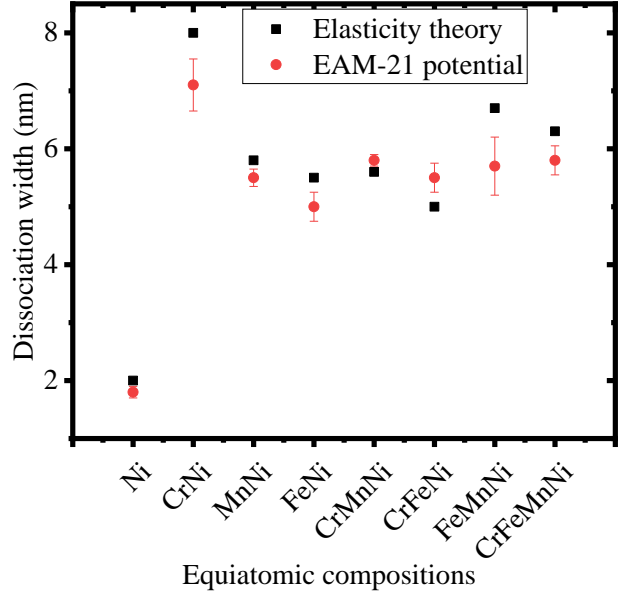

Figure 5: A comparison of the equilibrium dissociation distance evaluated with EAM-21 potential against elasticity theory at zero applied stress. The linear elasticity theory was reviewed in Hull and Bacon [28].

In figure 5, the focus was placed on the consistency of the equilibrium separation width predicted by EAM-21 potential compared to the elastic theory. The present calculation with EAM-21 potential exhibits consistency with elastic theory as shown in figure 5. Okamoto et al., [15] experimentally measured the dissociation width of CoCrFeMnNi to be 5.7±1 nm at 77 K with a weak-beam TEM, this measurement is consistent with dissociation width for equiatomic CrFeMnNi alloy predicted with EAM-21 potential which is given as 5.8±1.3 nm. In theory, the dissociation width is seen to be related to the value of $\gamma_{SFE}$. According to [28], it appears a low $\gamma_{SFE}$ causes large dissociation width. The equiatomic CrNi binary system has the lowest $\gamma_{SFE}$ (18 mJ/m$^2$), thus having the largest dissociation width ~7 nm compared to other Ni-based equiatomic alloys in Table 2.

## 3.6 *Evolution of the resolved shear stress*

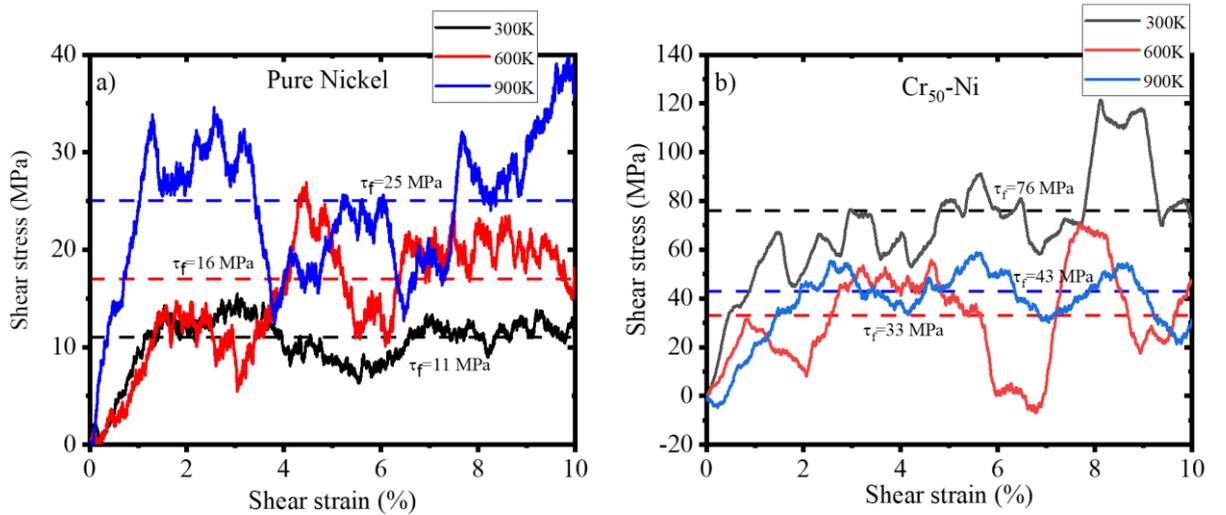

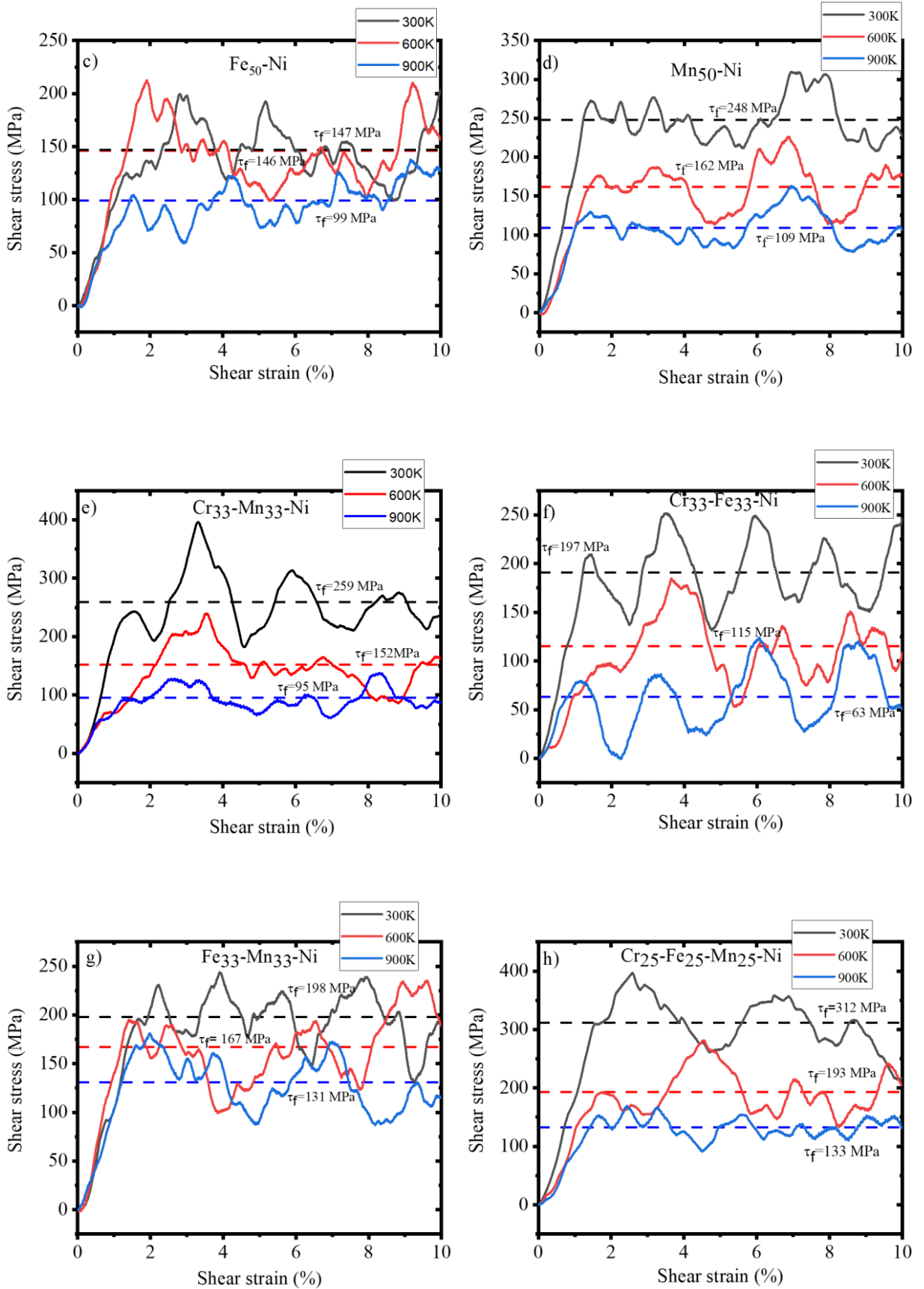

Figure 6: Evolution of the resolved shear stress (RSS) in a) Pure Nickel and b) Cr-Ni c) Fe-Ni d) Mn-Ni e) Cr-Mn-Ni f) Cr-Fe-Ni g) Fe-Mn-Ni h) Cr-Fe-Mn-Ni equiatomic alloy.

The dislocation motion was studied with an applied strain rate of $1 \times 10^8 \ s^{-1}$. In figure 6(a-h), the evolution of the resolved shear stress (RSS) is presented with a corresponding dislocation density and velocity of $0.7 \times 10^{14} \ m^{-2}$ and ~$45.21 \ m/s$, respectively. The black, red, and blue line indicate the average of instantaneous shear stress and $\tau_f$ is the friction stress defined as the average of the recorded shear stress. In all the alloys, the width of the stacking fault ribbon was seen to vary at $\pm 20$ % from the initial Shockley partial distance value during simulation time. The distribution of chemical elements in the system contributes to the wavy nature of the dislocation line, this is consistent with the works of Varvenne et al., [65] and Pasianot et al., [66]. Similar variations were observed experimentally for Cantor alloy by Otto et al., [67] using high-resolution transmission electron microscopy and Kawamura et al., [68] using the weak-beam technique in TEM. It is worth mentioning that in all cases, the crystal conserves fcc structure up to the maximum stress. The physical significance of this observation is consistent with the validation of the EAM-21 potential assuming that the fcc phase should be stable even under deformation.

*3.7 Influence of temperature on friction stress and critical resolved shear stress*

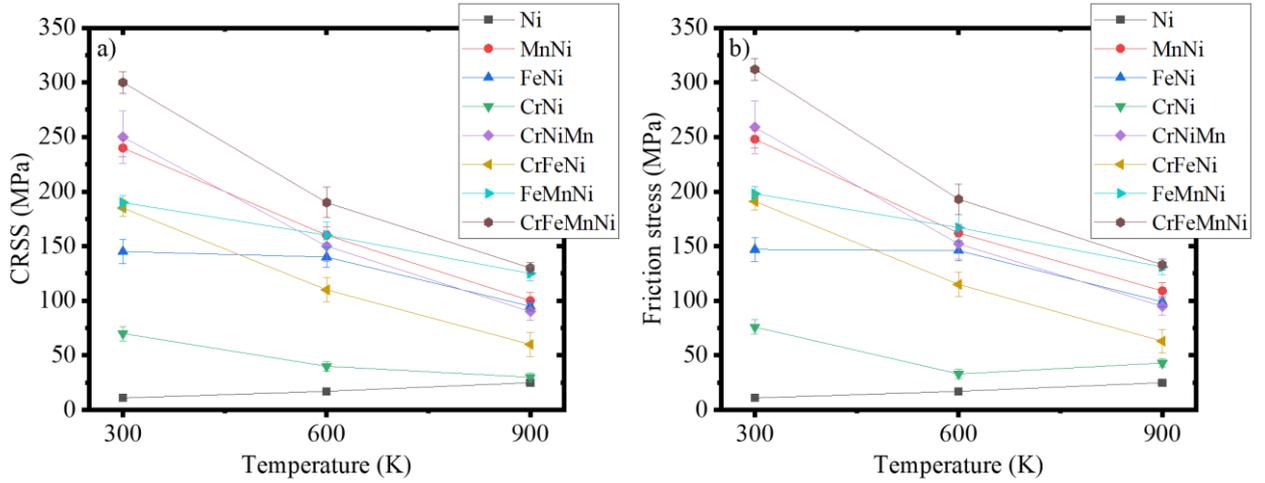

Figure 7: Variation of the friction stress and critical resolved shear stress (CRSS) as a function of temperature in FeNiCrMn equiatomic random alloy and its subsystems. a) friction stress and b) CRSS for dislocation motion in ½ [110] edge dislocation. Both friction stress and CRSS strongly depend on temperature. The error bar is the difference between minima and maxima of values obtained from three different seed numbers.

The determination of friction stress is necessary to model the mechanical behaviour and the associated properties of the materials. The ***friction stress*** $\tau_f$ is only a component of the ***critical resolved shear stress*** (CRSS) which is defined as the shear stress threshold for the single slip in the grains. If our results are to be compared to other single-phase materials, i.e., without precipitates or secondary phases, then the CRSS contains another component associated with pre-existing dislocations, the so-called forest strengthening $\tau_{forest}$. The latter has been found to obey the Taylor equation [69], i.e., $\tau_{forest} = \alpha \mu b \sqrt{\rho}$, where α is the forest strength, μ the shear modulus, and $b$ the norm of the Burgers vector. It is known that *α* decreases with the logarithm of dislocation density [70]. Since dislocation density is in the order of magnitude of $10^{13} \ m^{-2}$ before deformation, [71], one can deduce a value of approximately 20 MPa for the forest strengthening [72] and this component is known to be temperature and strain-rate independent. Inside the grain one can thus express: CRSS = 20 MPa + $\tau_f$. Using a Taylor model for the flow of polycrystals (with a Taylor factor M ≈ 3), one can finally express the yield stress of the material σ_y as:

$$\sigma_y = 60 + M\tau_f + \frac{K}{\sqrt{d}} \qquad (6)$$

where K is the Hall-Petch constant, and *d* is the average grain size. With a value of K ≈ 0.5 MPa.m$^{-0.5}$ given by Otto et al., [67] who reported also quasistatic straining as σ$_y$ ≈ 170 MPa at room temperature for polycrystal of large grains (negligible contribution of the Hall-Patch effect). This value implies thus that $\tau_f$ must be close to 37 MPa in quasi-static strain, which is one order of magnitude larger than $\tau_f$ found in our simulations (close to 320 MPa) as shown in figure 7. This discrepancy is due to the difference in strain rates between tensile tests and MD simulations. To obtain an order of magnitude of the effective strain rate in MD simulations, one can use the Orowan relation giving the shear rate $\dot{\varepsilon} = \dot{\gamma}/M = \rho b v/M$, where *v* is the dislocation velocity. Given our simulations conditions (ρ = $0.7 \times 10^{14}\ m^{-2}$ and $v = 45\ m/s$), one obtains the strain rate $\dot{\varepsilon}$ in the order of $2 \times 10^5\ s^{-1}$, which excludes comparison with an estimated quasi-static experimental tensile test. At high strain rates 4700 $s^{-1}$, Park et al., [73] reported yield stress close to 700 MPa for 16 μm grain-size polycrystals. The corresponding Hall-Petch effect is estimated to be 125 MPa. Using equation 6, one can deduce friction stress close to 172 MPa, which is now much closer to $\tau_f$ determined in MD simulations. A possible explanation of the remaining difference is probably that the strain rate in our MD simulations is still much larger than 4700 $s^{-1}$ used in experiments. It is expected that the friction will decrease with lower MD strain rates i.e., $1 \times 10^6\ s^{-1}$ - $1 \times 10^7\ s^{-1}$. All calculated values for friction stress can be seen in supplementary Table S.10.

### 3.8  *Influence of configurational entropy on friction stress*

The configurational entropy of mixing with randomly distributed atoms can enhance the mechanical properties of a material [74]. The regular solid solution model yields the relationship between the configurational entropy and chemical composition as:

$$\Delta S_{conf} = -k_B \sum_{i=1}^{n} x_i \log x_i \qquad (7)$$

where *n* is the number of elements, $x_i$ is the concentration of each element *i*, and $k_B$ is the Boltzmann constant. The configuration entropy $\Delta S_{conf}$ of all the binary, the ternary and the quaternary equiatomic alloy is estimated to be 0.69 k$_B$, 1.1 k$_B$, 1.39 k$_B$ respectively while the $\Delta S_{conf}$ for a pure element is equal to zero.

Yeh et al., [74] have hypothesised an increase in lattice strain is relative to the number of component elements present in a system. However, according to Wu et al., [75], He et al., [76], and Zhao et al., [77], this statement is questionable. For instance, in [78-79], the CoCrFeMnNi has the highest number of component elements in equiatomic compositions, but FeMnNi which is a ternary alloy has the largest lattice strain. One can deduce that alloying elements can either lead to an increase or decrease of the lattice strain of the system depending on the element's bonds with its nearest neighbours and the atomic size of the elements present in the system.

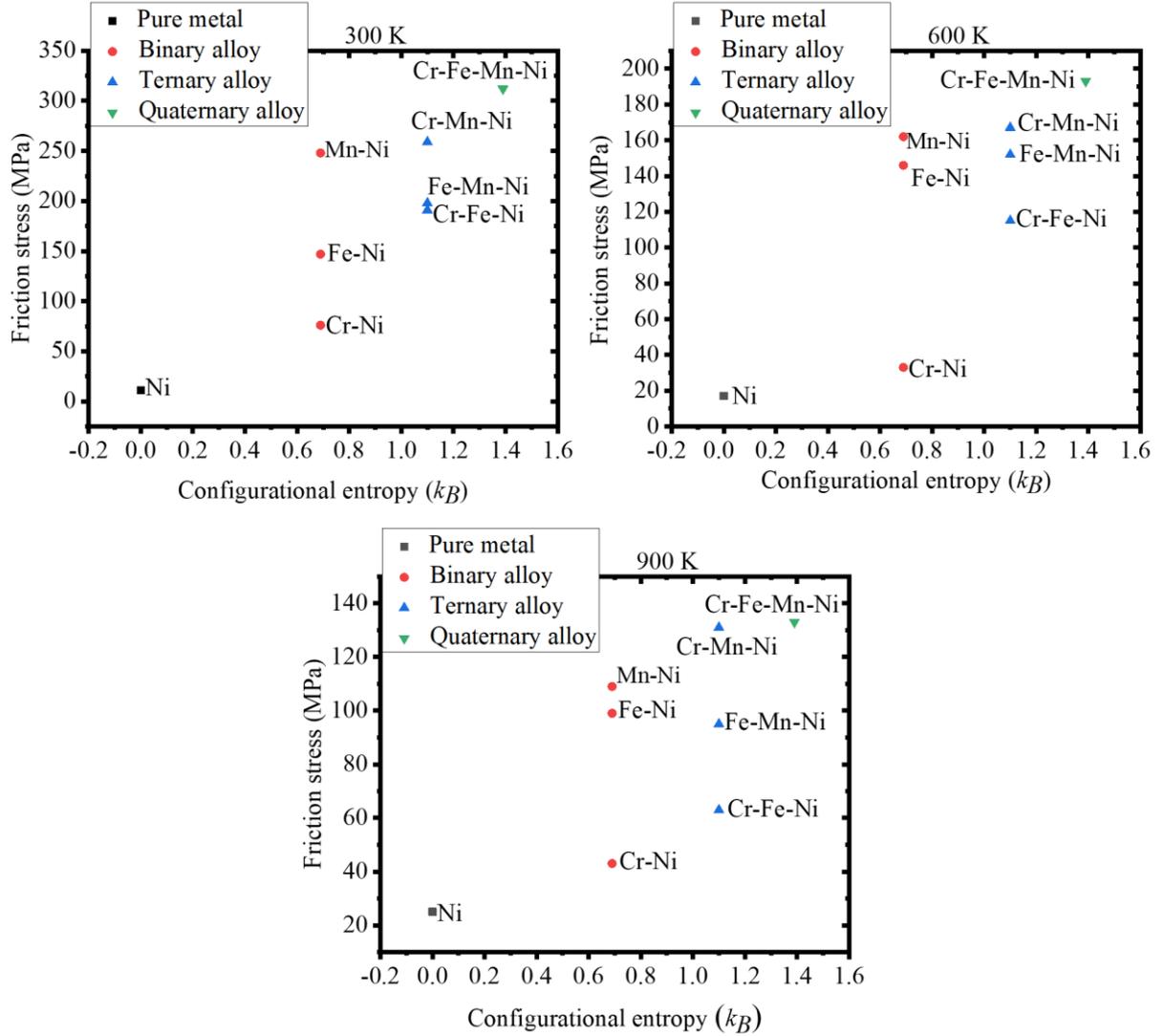

Figure 8: Friction stress as a function of configurational entropy in Ni-based equiatomic alloy (pure metal, binary alloy, ternary alloy, and quaternary alloy) at temperature range 300-900 K.

The friction stress is controlled by the interaction with different lattice strains induced by different atoms. Therefore, the friction stress can be expressed as a function of configurational entropy, as shown in figure 8. It is seen that the friction stress increases with configurational entropy. This means that dislocation moves more easily in less chemical complexity systems such as pure elements. Dislocation movements are not easy in multi-component alloy matrices such as ternary and quaternary systems since higher stress is needed to unpin their lattice strain to overcome the resistance of dislocation motion. According to figure 8, quaternary equimolar CrFeMnNi have the largest configurational entropy (larger alloy matrix), therefore the highest friction stress in the present Ni-based equiatomic alloy system. Let's note that this statement is proposed for the equimolar system and may not necessarily be conserved for the non-equimolar system, since $\Delta S_{conf}$ depends on the concentration of each element. This result is consistent with Wu et al., [75] and Zhao et al., [77]. It is also important to note that in figure 8, the system with the large friction stress is not necessarily the one with the highest number of elements. The alloying elements are also important as mentioned before, with the Mn-containing alloys mainly having the highest friction stress.

## 4  Conclusion

To investigate the mechanical properties of the CrFeMnNi system, an EAM potential named EAM-21 has been developed, by fitting to target values from experiments, DFT and CALPHAD based thermodynamic calculations. For CrFeMnNi alloys and their subsystems, the EAM-21 potential demonstrates a good agreement with available data such as elastic constants and stacking fault energy. The stability of the fcc phase under deformation and temperature conditions was confirmed.

Based on the good reproduction of the physical properties, preliminary simulations were carried out to validate the applicability of the EAM-21 potential for the investigation of plasticity mechanisms. Satisfactory conditions of glide of a dissociated ½<110>{111} edge dislocation could be obtained (based on MD simulations): the resulting evolution resolved shear stress (RSS) showed the motion of two partials connected by the stacking fault ribbon. The friction stress and CRSS of the glide of CrFeMnNi and its subsystems were also found to depend on temperature, in agreement with experiments.

These results suggest that the newly developed EAM-21 potential is suitable for the study of plasticity mechanisms in concentrated alloys from the CrFeMnNi family through atomistic simulation approaches.

## 5  Data availability

All the data that support the findings of this study are available from the corresponding authors (ayobami.daramola@emse.fr and fraczkie@emse.fr) upon reasonable request. The EAM-21 potential file (LAMMPS-readable) will be published on https://www.ctcms.nist.gov/potentials/ upon the publication of this manuscript.

### CRediT authorship contribution statement

**Ayobami Daramola**: Conceptualization, Methodology, Software, Validation, Formal analysis, Writing – original draft. **Giovanni Bonny**: Conceptualization, Methodology, Validation, Writing – review & editing, Supervision. **Gilles Adjanor**: Writing – review & editing, Supervision. **Christophe Domain**: Writing – review & editing, Supervision. **Ghiath Monnet**: Writing – review & editing, Supervision. **Anna Fraczkiewicz**: Conceptualization, Writing – review & editing, Project administration, Supervision, Funding acquisition.

### Declaration of competing interest

The authors declare that they have no known competing financial interests or personal relationships that could have appeared to influence the work reported in this work.

### Acknowledgements

This work is funded under the French ANR-PRCE-HERIA project (ANR-19-CE08-0012-01). One of the authors (GB) acknowledges funding from the Euratom research and training program 2014–2018 under grant agreement No. 755269 (GEMMA project).

# Appendix I

The optimized parameters set in equation (3) are summarized in Table A.1.

Table A.1: fitting parameters set for our potential, the unit of the distance and energy were measured in Angstrom and electron-Volt, respectively.

| $k$ | $r_k$ (Å) | $a_k (eV Å^{-3})$ |
|---|---|---|
| **FeMn** | | |
| 1 | 2.5 | 1.0147908536452896 |
| 2 | 2.9285714285714284 | 1.4257709924873434 |
| 3 | 3.3571428571428572 | -0.63479339435731807 |
| 4 | 3.7857142857142856 | 0.41754084083254533 |
| 5 | 4.2142857142857144 | -9.7696242824649351E-002 |
| 6 | 4.6428571428571423 | 9.5852666109606018E-003 |
| 7 | 5.0714285714285712 | -8.1325497863528454E-002 |
| 8 | 5.5 | 3.3311009576331700E-002 |
| **NiMn** | | |
| 1 | 2.0 | 95.006948162922356 |
| 2 | 2.2999999999999998 | 1.1071548091772379 |
| 3 | 2.5999999999999996 | 0.67688983543844305 |
| 4 | 2.8999999999999999 | 1.3516588206061959 |
| 5 | 3.1999999999999997 | 3.7150972237267252E-002 |
| 6 | 3.4999999999999996 | -0.69501999008087945 |
| 7 | 3.7999999999999998 | 0.88892722741483610 |
| 8 | 4.0999999999999996 | -0.40870992912638693 |
| **CrMn** | | |
| 1 | 2.0 | 97.463580469856765 |
| 2 | 2.2999999999999998 | 0.82448472122677030 |
| 3 | 2.5999999999999996 | 1.5434828988134552 |
| 4 | 2.8999999999999999 | -0.47969184617060157 |
| 5 | 3.1999999999999997 | 2.3150848233009671 |
| 6 | 3.4999999999999996 | -2.3452560010026682 |
| 7 | 3.7999999999999998 | 1.7977076422426939 |
| 8 | 4.0999999999999996 | -0.65925552837099599 |